\documentclass[10pt,twocolumn,letterpaper]{article}
\usepackage[accsupp]{axessibility} 
\usepackage{cvpr}              % To produce the CAMERA-READY version
\usepackage[dvipsnames]{xcolor}

\definecolor{cvprblue}{rgb}{0.21,0.49,0.74}
\usepackage[pagebackref,breaklinks,colorlinks,citecolor=cvprblue]{hyperref}
\usepackage{amsmath}
\usepackage{amssymb}
\usepackage{booktabs}
\usepackage{multirow}

\usepackage{textcomp}
\usepackage{tablefootnote}
\usepackage[accsupp]{axessibility}
\usepackage{hyperref}
\usepackage{graphicx}
\usepackage{marvosym}

\usepackage{bbding}
\hypersetup{pagebackref,breaklinks,colorlinks}
\newcommand{\ieno}{\textit{i.e.}}
\newcommand{\egno}{\textit{e.g.}}
\def\confName{CVPR}
\def\confYear{2024}

\title{NTIRE 2024 Challenge on Short-form UGC Video  Quality Assessment:\\ Methods and Results}

\author{Xin Li$^{\dag}$ \quad \quad Kun Yuan$^{\dag}$ \quad \quad  Yajing Pei$^{\dag}$ \quad \quad Yiting Lu$^{\dag}$ \quad \quad 
Ming Sun$^{\dag}$ \quad \quad Chao Zhou$^{\dag}$ \\ Zhibo Chen$^{\dag}$ \quad \quad Radu Timofte$^{\dag}$ \quad  \quad
Wei Sun\quad \quad Haoning Wu \quad \quad Zicheng Zhang \\Jun Jia \quad\quad  Zhichao Zhang\quad \quad Linhan Cao\quad \quad Qiubo Chen \quad \quad  Xiongkuo Min \quad \quad Weisi Lin \\ Guangtao Zhai\quad \quad  JianHui Sun\quad \quad  Tianyi Wang  \quad \quad Lei Li \quad \quad Han Kong \quad \quad Wenxuan Wang \\  Bing Li \quad \quad Cheng Luo \quad \quad Haiqiang Wang \quad \quad  Xiangguang Chen\quad \quad Wenhui Meng \\ Xiang Pan \quad \quad Huiying Shi \quad \quad Han Zhu \quad \quad  Xiaozhong Xu\quad \quad Lei Sun  \quad \quad Zhenzhong Chen \\   Shan Liu \quad \quad  Fangyuan Kong \quad \quad  Haotian Fan \quad \quad Yifang Xu \quad \quad  Haoran Xu \quad \quad  Mengduo Yang \\  Jie Zhou\quad \quad   Jiaze Li\quad \quad  Shijie Wen \quad \quad Mai Xu \quad \quad  Da Li \quad \quad   Shunyu Yao \quad \quad  Jiazhi Du \\  Wangmeng Zuo\quad \quad  Zhibo Li\quad \quad  Shuai He \quad \quad  Anlong Ming \quad \quad   Huiyuan Fu \quad \quad Huadong Ma \\ Yong Wu\quad \quad  Fie Xue\quad \quad  Guozhi Zhao \quad\quad  Lina Du \quad \quad  Jie Guo \quad \quad  Yu Zhang \\  Huimin Zheng \quad \quad  Junhao Chen\quad \quad  Yue Liu \quad\quad  Dulan Zhou \quad \quad   Kele Xu \\ \quad \quad  Qisheng Xu\quad \quad  Tao Sun\quad \quad  Zhixiang Ding\quad \quad  Yuhang Hu
}

\begin{document}
\thispagestyle{empty}
\twocolumn[{
\renewcommand\twocolumn[1][]{#1}% 
\maketitle
% \begin{center}
% \captionsetup{type=figure}
%     \setlength{\abovecaptionskip}{-0.001cm} 
%     \includegraphics[width=1\textwidth]{sec/Fig//intro.pdf}
%     \captionof{figure}{The two primary challenges of short-form videos: \textit{the kaleidoscope content} with various creation modes (top) and \textit{complicated distortion} arising from sophisticated video processing workflows (bottom). Regions with distortions are indicated by red boxes.
%     % The visualization of distorted regions is indicated by red boxes.
%     }
%     \label{fig:first}
% \end{center}
}] 

\maketitle
\renewcommand{\thefootnote}{}
\footnotetext{$^{\dag}$ X. Li (lixin666@mail.ustc.edu.cn), K. Yuan (yuankun03@kuaishou\\.com), Y. Pei, Y. Lu, M. Sun, C. Zhou, Z. Chen and R. Timofte are the challenge organizers.}

\footnotetext{The other authors are participants of the NTIRE 2024 Short-form UGC Video Quality Assessment Challenge.}

\footnotetext{The NTIRE2024 website: \textcolor{magenta}{https://cvlai.net/ntire/2024/}}

\footnotetext{The KVQ database: \textcolor{magenta}{https://lixinustc.github.io/projects/KVQ/}}

\renewcommand{\thefootnote}{\arabic{footnote}}

\begin{abstract}
This paper reviews the NTIRE 2024 Challenge on Short-form UGC Video Quality Assessment (S-UGC VQA), where various excellent solutions are submitted and evaluated on the collected dataset KVQ from popular short-form video platform, \textit{i.e.,} Kuaishou/Kwai Platform. The KVQ database is divided into three parts, including 2926 videos for training, 420 videos for validation, and 854 videos for testing. 
The purpose is to build new benchmarks and advance the development of S-UGC VQA. 
The competition had 200 participants and 13 teams submitted valid solutions for the final testing phase. The proposed solutions achieved state-of-the-art performances for S-UGC VQA. The project can be found at  ~\url{https://github.com/lixinustc/KVQ-Challenge-CVPR-NTIRE2024}.
\end{abstract}

\section{Introduction}
Short-from UGC video platforms, \eg, Kwai, and Tiktok have attained significant success and widespread popularity, attracting billions of users globally. In contrast to long-form UGC videos, short-form UGC (S-UGC) videos present several advantages, including mobile-friendly broadcasting, user-friendly engagement, kaleidoscope content creation, and snackable content, etc.~\cite{lu2024kvq}. However, S-UGC videos inevitably suffer from inconsistent and even poor subjective quality due to unprofessional creation, improper collection environments, or limited processing workflow. Building the quality assessment benchmarks tailed for S-UGC videos is crucial and urgent to advance the fast development of associated video processing techniques and quality control mechanisms for user-uploaded S-UGC videos. 

Recently, amounts of studies~\cite{ADAQA, BVQA, GSTVQA, VSFA, starVQA,discovqa, Dover,lu2024AIGC-VQA,yu2024video,li2023freqalign,liu2022liqa,lu2022styleam,DBLP:conf/mm/YuanKZSW23,DBLP:conf/mm/LiuWYSTZWL23,DBLP:conf/cvpr/ZhaoYSW23,DBLP:conf/cvpr/ZhaoYSLW23} have been proposed to assess the perceptual quality of long UGC videos or images by excavating the capability of pre-trained backbones, \ieno, ResNet~\cite{resnet,liu2022sourceBIQA,VSFA}, and Transformer~\cite{transformer,3dswin,liu2022swiniqa,lu2022rtn,simpleVQA,wu2022fastFastVQA,Yu2024ntire}. Since the scarcity of the VQA database, some well-designed sampling strategies~\cite{wu2022fastFastVQA,Dover}, \egno, fragment sampling, are introduced to reduce the computational complexity of the VQA method and enhance the diversity of VQA database. Apart from that, there are also some works exploring the potential of large cross-modality foundation models~\cite{ADAQA,maxwell,wu2023qQ-Align,CLIP} for VQA. However, the above works are only designed and validated on general long UGC VQA databases, \eg, LSVQ~\cite{ying2021patchLSVQ}, YouTube-UGC~\cite{youtubeUGC}, and KoNViD-1k~\cite{KoNViD-1k}, lacking the perception capability for short-form UGC VQA. 

This NTIRE 2024 Short-form UGC Video Quality Assessment Challenge is held to promote related research on S-UGC VQA, and develop the powerful S-UGC VQA benchmark to assist the technique evolution of S-UGC video creation, compression, or processing, etc. In this competition, we utilize the KVQ dataset~\cite{lu2024kvq} collected from the Kwai platform to evaluate the submitted methods. The KVQ dataset contains 4200 S-UGC videos with distinct creation modes, content scenarios, and video processing workflows, satisfying the data distribution of practical short-form video platforms. For evaluation, the KVQ dataset is divided into three parts with a proportion of 7:1:2 for training, validation, and testing. Instead of only quality scores, we also provide the ranking labels for indistinguishable samples in the validation and testing parts, ensuring the fine-grained perception evaluation for S-UGC VQA methods. The evaluation metric is composed of four parts, including PLCC, SROCC, the ranking accuracies between homogeneous video pairs (\ieno, with the same content) and non-homogeneous video pairs.  \textit{Notably, we have released the quality labels and ranking scores of the validation data, and reopened the test submission on Codalab~\cite{codalab_competitions_JMLR} after this competition, intending to facilitate the development of algorithms.} 

The competition has two phases, \ie, the development and testing phases, attracting 200 registered participants in total. There are 49 teams, and 56 teams submitting their predictions in the development and testing phases, respectively. Finally, 13 teams submitted their fact sheets and validation codes for competition ranking. Their algorithms will be summarized in the Section~\ref{sec:teams_methods}.  

This challenge is one of the NTIRE 2024 Workshop~\footnote{https://cvlai.net/ntire/2024/} associated challenges on: dense and non-homogeneous dehazing~\cite{ntire2024dehazing}, night photography rendering~\cite{ntire2024night}, blind compressed image enhancement~\cite{ntire2024compressed}, shadow removal~\cite{ntire2024shadow}, efficient super resolution~\cite{ntire2024efficientsr}, image super resolution ($\times$4)~\cite{ntire2024srx4}, light field image super-resolution~\cite{ntire2024lightfield}, stereo image super-resolution~\cite{ntire2024stereosr}, HR depth from images of specular and transparent surfaces~\cite{ntire2024depth}, bracketing image restoration and enhancement~\cite{ntire2024bracketing}, portrait quality assessment~\cite{ntire2024QA_portrait}, quality assessment for AI-generated content~\cite{ntire2024QA_AI}, restore any image model (RAIM) in the wild~\cite{ntire2024raim}, RAW image super-resolution~\cite{ntire2024rawsr}, short-form UGC video quality assessment~\cite{ntire2024QA_UGC}, low light enhancement~\cite{ntire2024lowlight}, and RAW burst alignment and ISP challenge.

% This challenge is one of the NTIRE 2024 Workshop\footnote{https://cvlai.net/ntire/2024/} challenges:   

% \textit{Notably, the quality labels and ranking results have been released after competition, and the }

\section{Challenge}
\label{sec:challenge}
The NTIRE 2024 Short-form UGC Video Quality Assessment Challenge is organized to advance the development of VQA techniques for short-form UGC videos. Moreover, this is the first challenge focusing on S-UGC VQA, aiming to build a new benchmark to guide the perceptual quality improvement of S-UGC videos. The details of the whole challenge are described in the following parts, including datasets, evaluation protocol, and challenge phases.

\begin{table*}[htp]
\caption{Quantitative results of the NTIRE 2024 Short-form UGC Video Quality Assessment Challenge. }
\centering
\resizebox{1.0\textwidth}{!}{\begin{tabular}{ccc|c|cccc|c|c|c}
\toprule
\multicolumn{1}{c|}{Rank} & Team & Leader & Final Score & SROCC & PLCC & Rank1 & Rank1 & Pretrained &  Ensemble & Extra Data \\  \midrule
\multicolumn{1}{c|}{1}    &  SJTU MMLab    &  Wei Sun    & 0.9228  &  0.9361 &   0.9359   &  0.7792     &  0.8284     &  $\surd$          & $\surd$ & $\surd$ (public) \\
\multicolumn{1}{c|}{2}    &  IH-VQA    &   Jianhui Sun       &    0.9145  &  0.9298 & 0.9325     & 0.7013      &  0.8284     &   $\surd$         & $\surd$  & $\surd$ (private)  \\
\multicolumn{1}{c|}{3}    & TVQE     &   Haiqiang Wang       &    0.9120  & 0.9268  & 0.9312     & 0.6883      &   0.8284    &  $\surd$       &  $\surd$ & $\surd$ (private)  \\
\multicolumn{1}{c|}{4}    &  BDVQAGroup    &  Fangyuan Kong        &   0.9116   & 0.9275  &  0.9211    &     0.7489  &  0.8462     &            $\surd$      & $\surd$  & $\surd$ (public) \\
\multicolumn{1}{c|}{5}    &  VideoFusion    &   Haoran Xu       &  0.8932    &  0.9026 &  0.9071    & 0.7186      &  0.8580     &   $\surd$        &  $\surd$ & $\surd$ (public) \\
\multicolumn{1}{c|}{6}    &   MC$^2$Lab   &  Shijie Wen        &  0.8855    & 0.8966  &  0.8977    &  0.7100     &   0.8521    &  $\surd$         &  $\surd$ &  $\times$ \\
\multicolumn{1}{c|}{7}    &   Padding   &  Da Li       &  0.8690    &  0.8841 & 0.8839     &  0.6623     &  0.8047     &    $\times$        & $\times$ & $\times$ \\
\multicolumn{1}{c|}{8}    &   ysy0129   &  Shunyu Yao      &  0.8655    & 0.8759 & 0.8777     &   0.6883    &   0.8402    &   $\surd$         & $\surd$   & $\times$\\
\multicolumn{1}{c|}{9}  &   lizhibo   &  Zhibo Li      &  0.8641    &  0.8778 & 0.8822     & 0.6494      & 0.7929      &    $\surd$    &  $\times$ &$\times$\\
\multicolumn{1}{c|}{10} &   YongWu   &  Yong Wu      &  0.8555    &  0.8629 &  0.8668    &  0.6970    &  0.8462     & $\times$           & $\times$ & $\times$\\
\multicolumn{1}{c|}{11} &   we are a team   &  Lina Du, Jie Guo     & 0.8243     & 0.8387  & 0.8324     &    0.6234   &  0.8225     & 
 $\times$         & $\times$  &  $\times$\\
\multicolumn{1}{c|}{12} &   dulan   &  Dulan Zhou     & 0.8098     & 0.8164  &   0.8297   &   0.5758    &      0.8047 &    $\surd$       &  $\surd$   &$\times$ \\
\multicolumn{1}{c|}{13} &  D-H   &  Zhixiang Ding     & 0.7677     & 0.7774  &  0.7832    &   0.5931    & 0.7160      &        $\surd$    &$\times$  & $\times$ \\ \midrule \midrule

 \multicolumn{2}{c}{\multirow{3}{*}{Baseline}} &  VSFA~\cite{VSFA} &   0.7869 & 0.7974 & 0.7950 & 0.6190 & 0.7870 & \multirow{3}{*}{$\times$} & \multirow{3}{*}{$\times$} & \multirow{3}{*}{$\times$} \\

 & & SimpleVQA~\cite{simpleVQA} & 0.8159 & 0.8306 & 0.8202 & 0.6147 & 0.8461 & & & \\

 && FastVQA~\cite{wu2022fastFastVQA} & 0.8356  & 0.8473 & 0.8467 & 0.6494 & 0.8166 &&& \\ \bottomrule
 
\end{tabular}}
\label{tab:results}
\end{table*}

\subsection{Dataset}
\label{sec:dataset}
To ensure the reliable and comprehensive evaluation for each method, we utilize our KVQ dataset~\cite{lu2024kvq} collected from the practical S-UGC video platform, \ie, Kwai/Kuaishou platform. The KVQ dataset contains 4200 S-UGC videos, consisting of 600 user-uploaded videos and 3600 processed videos via various practical processing workflows in Kuaishou platform. In the collection process, we select video samples from 9 content scenarios, \ie, food, stage, computer graphic, night, caption, person, crowd, landscape, and portrait, covering several typical creation modes, including three-stage, live, subtitled, and special effect videos, etc. 

In practical S-UGC video platforms, video processing tools are diverse and complicated. Consequently, we apply three typical processing techniques, including enhancement, pre-processing, and transcoding. Among them, enhancement tools consist of de-artifact, de-noise, and deblur, and pre-processing tools can be divided into global-level and ROI-level. The transcoding tools aim to reduce the bitstream, where the quantization parameters (QPs) from 16 to 47 are divided into six groups. The above three processing strategies are serially applied to user-uploaded S-UGC videos, resulting in various degradations. 

These S-UGC videos are annotated by 15 professional researchers specializing in image processing. The Mean Opinion Score (MOS) value is [1-5] and the scoring interval is 0.5, which makes it easy for humans to annotate the quality score. Notably, the fine-grained quality differences between different S-UGC video samples are hard to identify. In the validation and testing datasets of this challenge, we also provide the ranking score for the video samples with similar quality scores. In the test stage, we annotated the ranking scores for 169 homologous video pairs (\ie, with the same contents but different degradations), and 231 non-homologous video pairs, respectively. For the validation dataset, we provide ranking scores for 62 homologous video pairs and 38 non-homologous video pairs, respectively. The above hybrid annotations enable a more thorough evaluation of submitted VQA methods. More details of the KVQ dataset can be found in work~\cite{lu2024kvq}.

\subsection{Evaluation protocal}
\label{sec:evaluation_protocal}
In this challenge, we aim to evaluate each method from four perspectives: (i) the prediction monotonicity with Spearman rank-order correlation coefficient (SROCC) and (ii) the prediction accuracy with Person Linear Correlation coefficient (PLCC); (iii) fine-grained ranking accuracy between homologous video pairs, \ie, Rank1; (iv) the challenging ranking accuracy between non-homologous video pairs, \ie, Rank2. The final score used for ranking is computed by combining the above metrics:
\begin{align}
\centering
   \notag \textrm{Final\_Score} &= 0.45*\textrm{SROCC}+0.45*\textrm{PLCC}\\ 
    & +0.05*\textrm{Rank1}+0.05*\textrm{Rank2}
    \label{eq:final_score}
\end{align}
We expect the above equation can estimate both coarse- and fine-grained quality assessment capability for submitted S-UGC VQA methods.

\subsection{Challenge phases}
\label{sec:challenge_pahses}
There are two phases in this challenge, \ie, the development phase, and testing phase.

\noindent\textbf{Development phase:} In this phase, we divided our collected KVQ dataset into training, validation, and testing data based on their contents. We release 2926 S-UGC videos and their corresponding quality scores to participants, which aims to support them develop their S-UGC VQA algorithm. Besides, we provide 420 S-UGC videos without quality labels for validation. Participants can upload their submissions to the challenge platform and obtain their final score, SROCC, PLCC, Rank1, and Rank2 accuracies. We also release the benchmark code based on SimpleVQA~\cite{simpleVQA} to help participants quickly familiarize the process of S-UGC VQA. In the development phase, 881 submissions from 49 teams are received.   

\noindent\textbf{Testing phase}
In the test phase, we release 854 S-UGC videos for testing on the challenge platform. The testing leaderboard is hidden and each team can observe their testing results individually. The ranking for this challenge is based on the final score as Eq.~\ref{eq:final_score} in the testing stage. After testing, the participants are requested to submit the fact sheet and source code/executable to reproduce their results. There are 56 teams submitting their prediction results on the challenge platform. Finally, we received the fact sheets and source codes from 13 teams, which are used for final ranking.

\section{Challenge Results}
\label{sec:challenge_results}
The challenge results are shown in Table~\ref{tab:results}. We only report the performances of the teams submitting their fact sheets. From this Table, we can find the top five teams, including SJTU MMLab, IH-VQA, TVQE, BDVQAGroup, and VideoFusion have achieved excellent results on both PLCC and SROCC, which exceed 0.9. Among all teams, Padding, lizhibo, YongWu, we are a team and D-H do not use ensemble strategies and extra data. MC$^2$Lab, ysy0129 and dulan do not utilize extra data. However, the above methods still achieve great performances in this competition.  

We also report the performances for existing baselines, including VSFA~\cite{VSFA}, SimpleVQA~\cite{simpleVQA}, and FastVQA~\cite{wu2022fastFastVQA}, without utilizing the ensemble and extra data. We can observe that 10 teams have achieved better performances than these baselines.
The methods of each team are described in Sec.~\ref{sec:teams_methods}, which greatly promote the development of short-form UGC video quality assessment.

\begin{figure*}[htp]
    \centering
\includegraphics[width=0.9\textwidth]{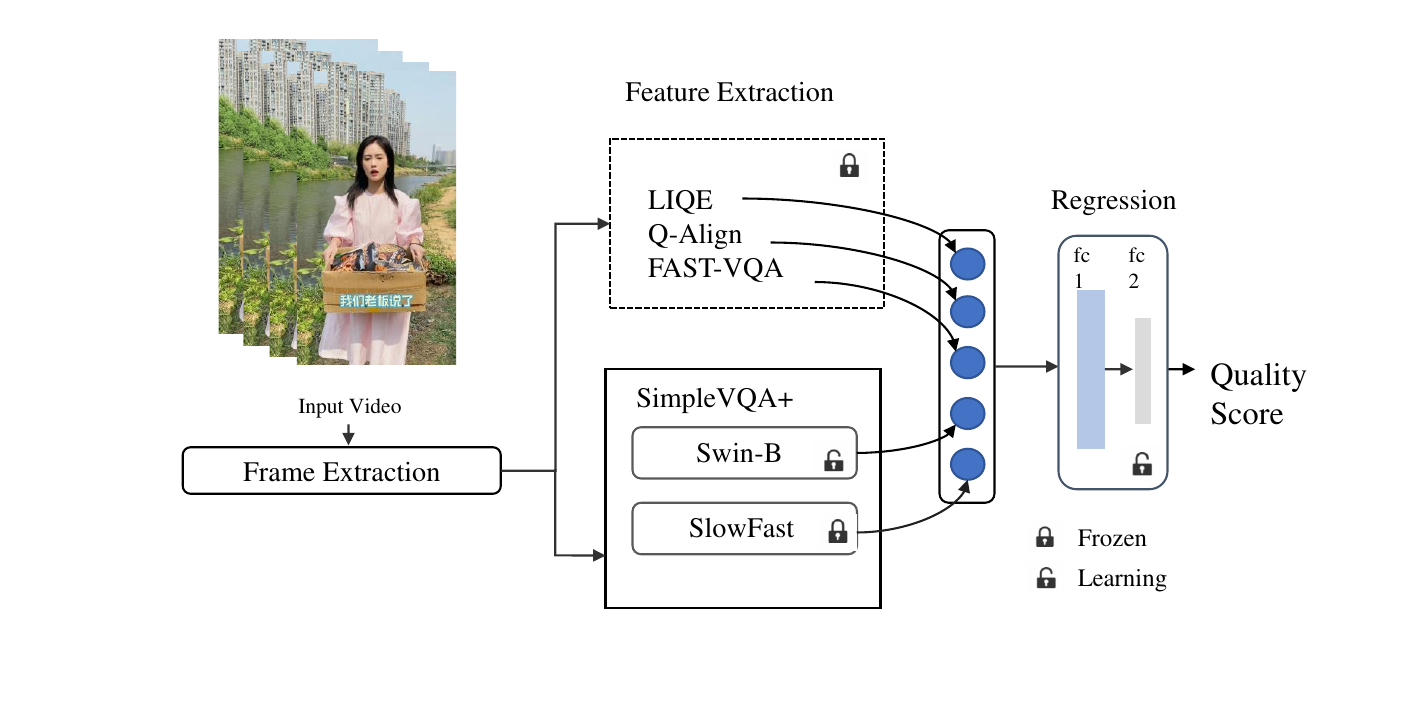}
    \caption{The network architecture~\cite{sun2024Enhancing} of the solution proposed by team SJTU MMLab.}
    \label{fig:sjtu_mmlab}
\end{figure*}

\section{Teams and Methods}
\label{sec:teams_methods}
\subsection{SJTU MMLab}
This team proposes the BVQA method, which is based on SimpleVQA+~\cite{simpleVQA, sun2023analysisSimpleVQA2}, comprising a Swin Transformer-B~\cite{3dswin} for spatial feature extraction and a temporal pathway of SlowFast for temporal feature extraction. The whole framework is shown in Fig.~\ref{fig:sjtu_mmlab}. Given the diverse visual content and complexity distortions in the KVQ dataset, they incorporate three BI/VQA models: LIQE~\cite{LIQE}, Q-Align~\cite{wu2023qQ-Align}, and FAST-VQA~\cite{wu2022fastFastVQA}, to extract comprehensive quality-aware features to aid their BVQA model. Specifically, LIQE extracts eleven types of distortions, nine scene categories, and five quality-level probabilities as quality-related and scene-specific features. Q-Align extracts quality-level features for video frames by the large multi-modality models, while FAST-VQA extracts spatial-temporal quality-aware features from local video patches. They concatenate these features with those from SimpleVQA+ and utilize a two-layer MLP to derive the video quality scores.

\noindent\textbf{Training Details.}
In the above BVQA method, the weights of SimpleVQA+ are initialized by training it on the LSVQ~\cite{ying2021patchLSVQ} dataset. LIQE is trained on LIVE~\cite{sheikh2006statisticalLIVE},
CSIQ~\cite{larson2010mostCSIQ}, KADID-10k~\cite{lin2019kadid-10k}, BID~\cite{ciancio2010noBID}, CLIVE~\cite{ghadiyaram2015massiveCLIVE}, and KonIQ-10k~\cite{hosu2020koniqKonIQ-10k}.
Q-Align is trained on SPAQ~\cite{fang2020perceptualSPAQ}, KonIQ-10k~\cite{hosu2020koniqKonIQ-10k}, KADID-10k~\cite{lin2019kadid-10k}, LSVQ~\cite{ying2021patchLSVQ}, and AVA~\cite{murray2012avaAVA}. FAST-VQA is also trained on LSVQ~\cite{ying2021patchLSVQ}. 
They sample one key frame from one-second video chunks (\ie, 1 fps) for the spatial feature extraction module of SimpleVQA+ as well as for extracting LIQE and Q-Align features. The resolution of key frames is further resized to 384$\times$384 for training. For the temporal feature extraction module, the resolution of the videos is resized to 224$\times$224. They then split the whole video into several one-second length video chunks to extract the corresponding temporal features. Fast-VQA features are extracted from the entire video using the fragment sampling method~\cite{wu2022fastFastVQA}. They train the proposed model on $2$ Nvidia RTX 3090 GPUs with a batch size $6$ for $30$ epoches. The learning rate is set as $1e-5$.

\noindent\textbf{Testing Details.}
They randomly split the publicly available data of
the KVQ dataset into ten different training-validation sets and train ten
BVQA model. The video quality score is computed by averaging the quality scores obtained from these models.

\subsection{IH-VQA}
\begin{figure*}[htp]
    \centering
    \includegraphics[width=0.9\textwidth]{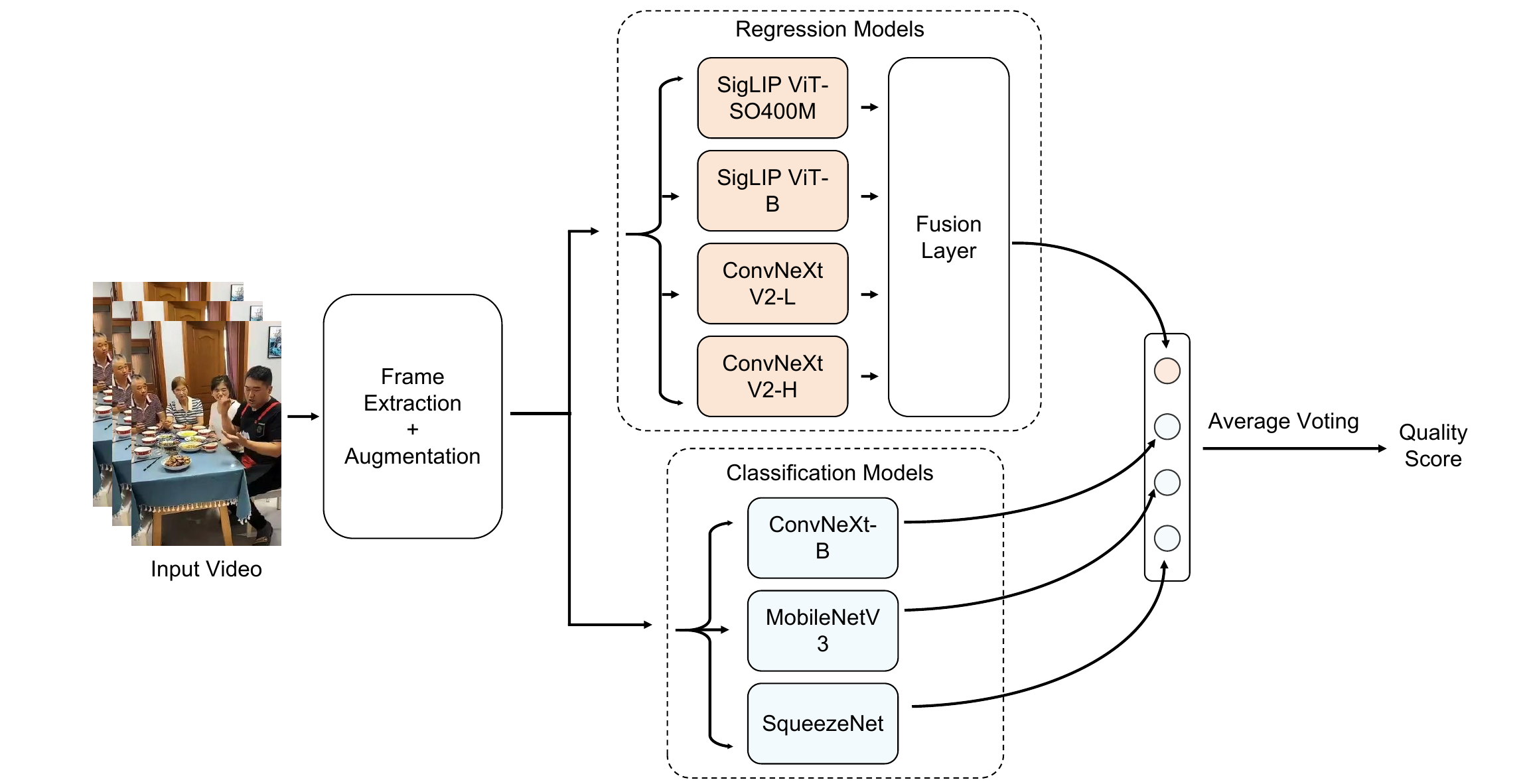}
    \caption{The framework of the Ensemble-based Video Quality Assessment System proposed by IH-VQA.}
    \label{fig:IH-VQA}
\end{figure*}

This team proposes an Ensemble-based Video Quality Assessment System, which consists of seven different expert models, \ie, four regression expert models, including SigLIP-ViT-SO400M~\cite{Zhai2023SigmoidLF}, SigLIP-ViT-B~\cite{Zhai2023SigmoidLF}, ConvNeXt V2-H~\cite{Woo2023ConvNeXtVC}, ConvNeXt V2-L~\cite{Woo2023ConvNeXtVC}, and three classification expert models, consisting of ConvNeXt-B~\cite{liu2022convnetConvNetXt}, MobileNetV3~\cite{Howard2019SearchingFM} and SqueezeNet~\cite{Iandola2016SqueezeNetAA}.
The system framework is outlined in Fig.~\ref{fig:IH-VQA}. As shown in this figure, four video frames are inputted into these four regression models, respectively, and their features are fused in the last pooling layer. After that, two additional feed-forward layers are used to produce the final prediction of the input video. They also adopt three classification models to predict the probability of being rated in different score intervals. These classification expert models rate the given video frames, and their votes are averaged as the final prediction.
To address the quality variance issues within different frames, they designed a novel loss function, \ie, the mean absolute error between the target quality label and the mean of the predictions from multiple frames, jointly with the cross-entropy loss.

\noindent\textbf{Training Details.}
Four regression models are pre-trained on an in-house dataset with 20K data and then fine-tuned on the official dataset KVQ. During training, techniques like the Fast Gradient Method and Exponential Moving Average (EMA)~\cite{Klinker2011ExponentialMAEMA} are also exploited for better performance. For three classification models,
the weights of their spatial attention blocks are initialized to zero.
The remaining model blocks keep the pre-trained weights on the ImageNet-1K dataset~\cite{Russakovsky2014ImageNetLS}.
They also use EMA to boost model performance. Once they get the well-trained ConvNeXt-B model, it will serve as a teacher model for distillation purposes, aiding in the training of two additional student models (\ie, MobileNetV3 and SqueezeNet).
Their models are trained for nine GPU hours with 20 epochs. An Adam optimizer with a learning rate of $3e^{-5}$ is leveraged to train large models like SigLIP and ConvNeXt, while an Adam optimizer with a learning rate of $1e^{-3}$ is used to train smaller models such as MobileNetV3 and SqueezeNet.

\noindent\textbf{Testing Details.}
During inference, the final prediction is determined by averaging the predictions of the above seven models.

\subsection{TVQE}
\begin{figure*}[htp]
    \centering
\includegraphics[width=0.98\textwidth]{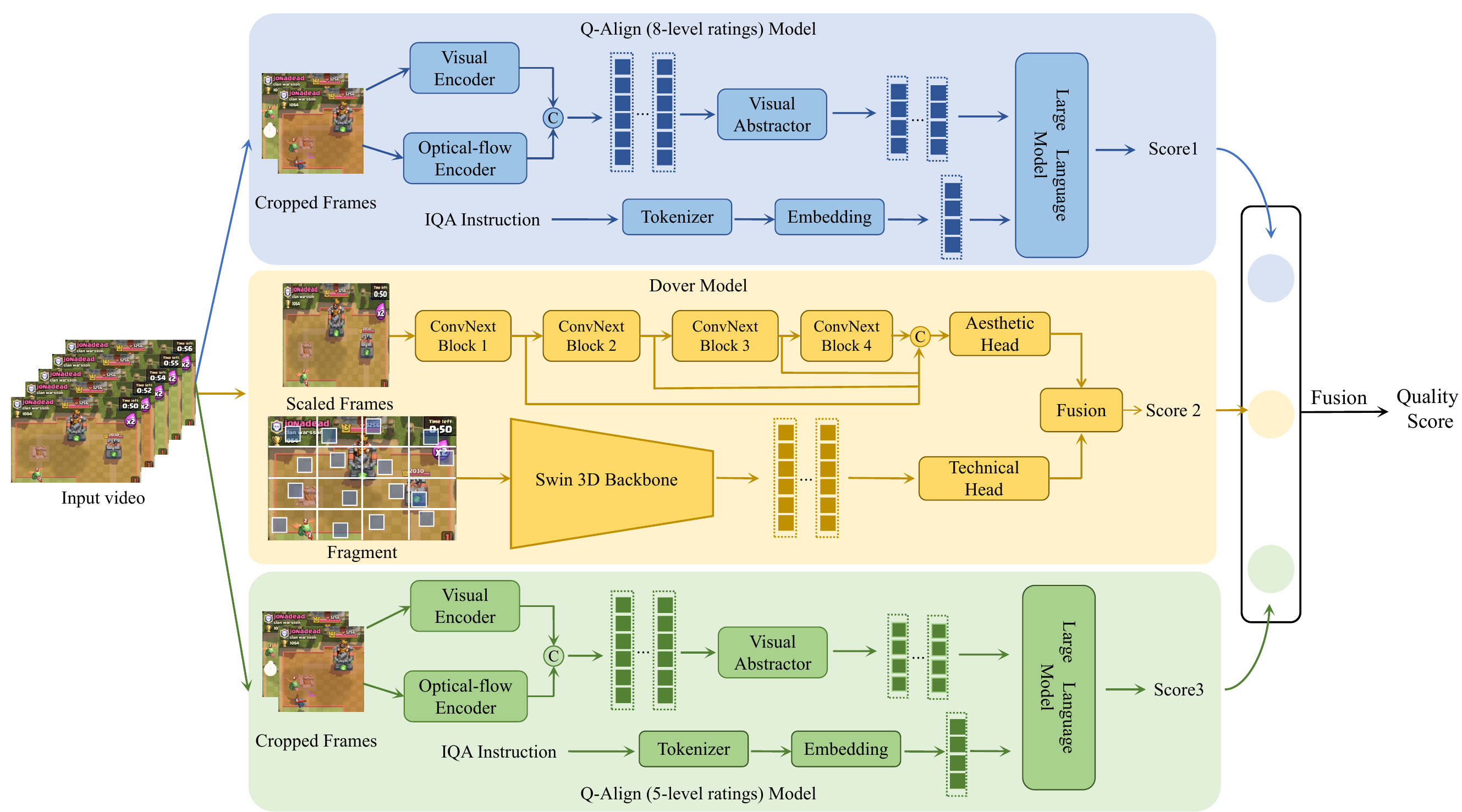}
    \caption{The network architecture of the solution proposed by team TVQE.}
    \label{fig:tvqe}
\end{figure*}

This team proposes the TVQE, which is a hybrid model trained for VQA tasks. The overall framework is shown in Fig.~\ref{fig:tvqe}. It combines two multi-modality models to extract visual information and semantic information and a classical convolution neural network to capture technical and aesthetic quality.

First, they introduce a feature pyramid aggregation mechanism on the backbone, \ie, the ConvNeXt~\cite{liu2022convnetConvNetXt}, of the aesthetics branch of Dover~\cite{wu2023exploringDover} to extract better aesthetic representations. This is motivated by the fact that the
aesthetics of short-form video is vital to the final quality.

They adopt two large multi-modality models, \ie, Q-align~\cite{wu2023qQ-Align}, with different
levels of quality descriptions. The aim is to alleviate the quantization effect around 5-level quality descriptors, which is introduced in the process to convert predicted rating levels to the final predicted score. Furthermore, as a dedicated video quality metric, they integrate the optical flow module
to capture motion information into the assessment process. The extracted motion feature is combined with a visual feature to seamlessly represent the quality of the entire video.
These three models were trained independently on the official KVQ dataset~\cite{lu2024kvq} and private datasets in an end-to-end manner. During the inference stage,
the final predicted score could be obtained by heuristically fusing the
prediction results of these models. 
\begin{figure}[htp]
    \centering
\includegraphics[width=1\linewidth]{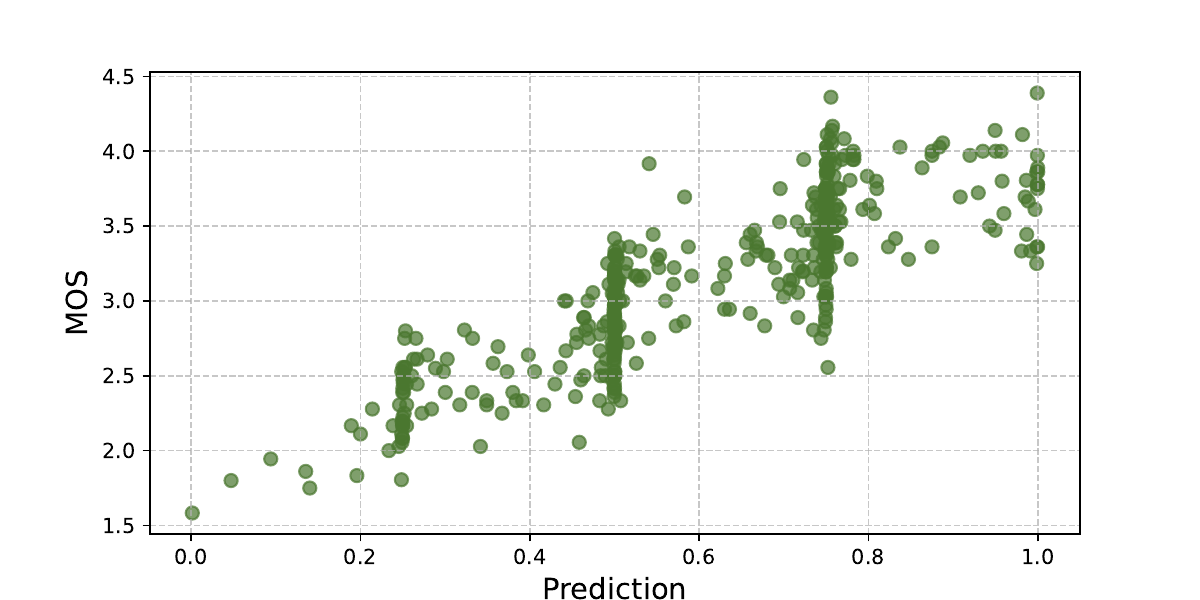}
    \caption{The quantization effect when the original Q-Align is tested on the split of the training dataset. The experimental results are from team TVQE.}
    \label{fig:quantization_effect}
\end{figure}

\noindent\textbf{Experimental Analysis.} They have tested several top-performing VQA methods in the validation stage. In general, two SOTA metrics, i.e. Dover~\cite{wu2023exploringDover} and Q-align~\cite{wu2023qQ-Align} are selected
due to their superior performance. Meanwhile, they have found the motion
feature is absent in the original Q-align. This might lead to sub-optimal
representation for the video quality assessment task. Furthermore, they
find a quantization effect in Fig.~\ref{fig:quantization_effect} when the original Q-align is evaluated on a split of the training set, even though the obtained PLCC and SRCC are not bad.

\noindent\textbf{Training Details.}
They train three models, \ie, Dover~\cite{wu2023exploringDover}, Q-Align~\cite{wu2023qQ-Align}, and IFRNet~\cite{kong2022ifrnet} on the above dataset independently. The training datasets consist of three parts, including the KVQ training part, a private UGC dataset, and a private PGC dataset. Among them, The private UGC dataset also contains short-form videos. The processing pipeline includes enhancement filters before transcoding. The private PGC dataset contains multiple distortion types, such as transcoding with different codecs, enhancement followed by
transcoding. 
These three 
datasets are simply concatenated together.

\noindent\textbf{Testing Details.}
A test video is processed by the three models independently. The predicted
scores are heuristically fused together as the final prediction.

\subsection{BDVQAGroup}
\begin{figure*}[htp]
    \centering
\includegraphics[width=0.98\textwidth]{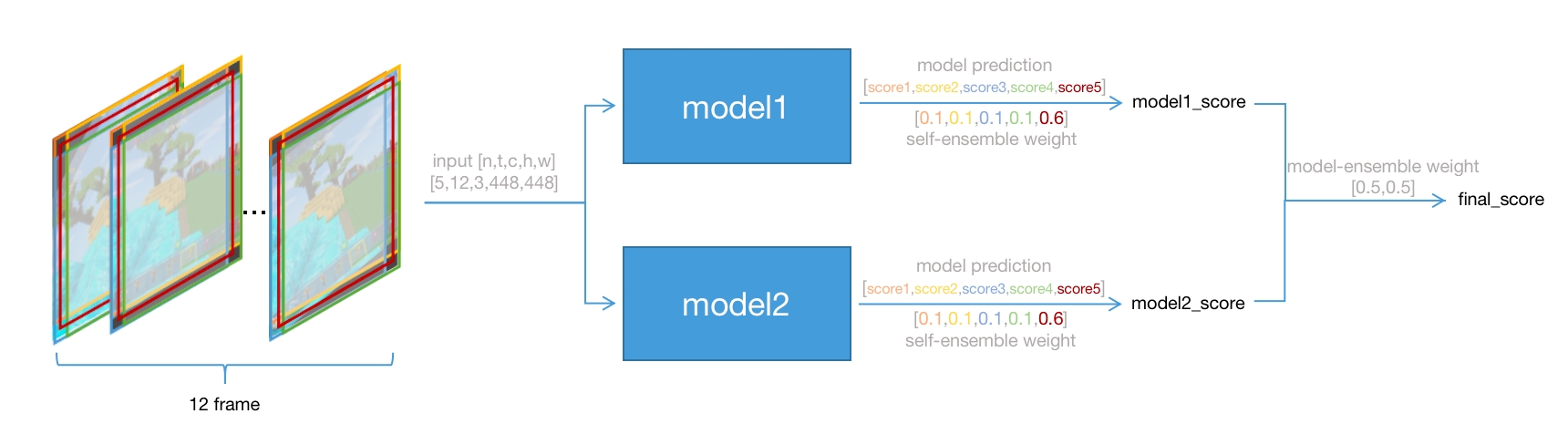}
    \caption{The network architecture of the solution proposed by team BDVQAGroup.}
    \label{fig:bdvqagroup}
\end{figure*}

 This team chooses a method called Q-Align to assess the quality of UGC videos. It is based on large multi-modality models (LMMs). Q-Align converts MOS scores into rating levels, and uses classification to teach LMMs with text-defined rating levels instead of scores. During inference, it extracts the close-set probabilities of rating levels and performs a weighted average to obtain the LMM-predicted score. They raise serval training and inference tricks to increase the performance of this method. The whole framework is depicted in Fig.~\ref{fig:bdvqagroup}. 

 \noindent\textbf{Training Details.} 
This team uses several data augmentation methods to increase
the training dataset and enhance the robustness of the model, which are listed as follows:
\begin{itemize}
    \item data resample: calculate the maximum and minimum mos score of
training dataset, divide them into 5 intervals for rating level mapping and then resample offline until each rating level has the same amount
of training data.
\item random frame sample: during training, 8 frames are randomly selected from a video as input.
\item random reverse: video frames are randomly arranged in chronological or reverse order.
\item random crop: randomly crop frames at a ratio of 0.9, and the frames of the same video crop at the same position.
\end{itemize}

This team uses a Q-Align~\cite{wu2023qQ-Align} model pre-trained on KonIQ~\cite{hosu2020koniqKonIQ-10k}, SPAQ~\cite{fang2020perceptualSPAQ}, KADID~\cite{lin2019kadid-10k}, AVA~\cite{murray2012avaAVA},
and LSVQ~\cite{ying2021patchLSVQ}, and fine-tune this model through two strategies. The first
model is fine-tuned for 3 epochs on the NTIRE 2024 KVQ dataset. The second model
is fine-tuned for 1 epoch on the ICME 2021 UGCVQA~\cite{ICME2021UGC-VQA}, and then fine-tuned for
another 3 epochs on the NTIRE 2024 KVQ dataset.

\noindent\textbf{Testing Details.}
The following ensemble methods are implemented to increase
model performance. 
\begin{itemize}
\item self-ensemble: during inference, 12 frames are taken from each video
at equal intervals, and FiveCrop is performed on each frame.
Because of the sandwich video, they give higher weight to the prediction
of the central area.
\item model ensembles: they use two models to predict and give each prediction the same weight to get the final result.
\end{itemize}

\subsection{VideoFusion}
\begin{figure*}[htp]
    \centering
\includegraphics[width=0.98\textwidth]{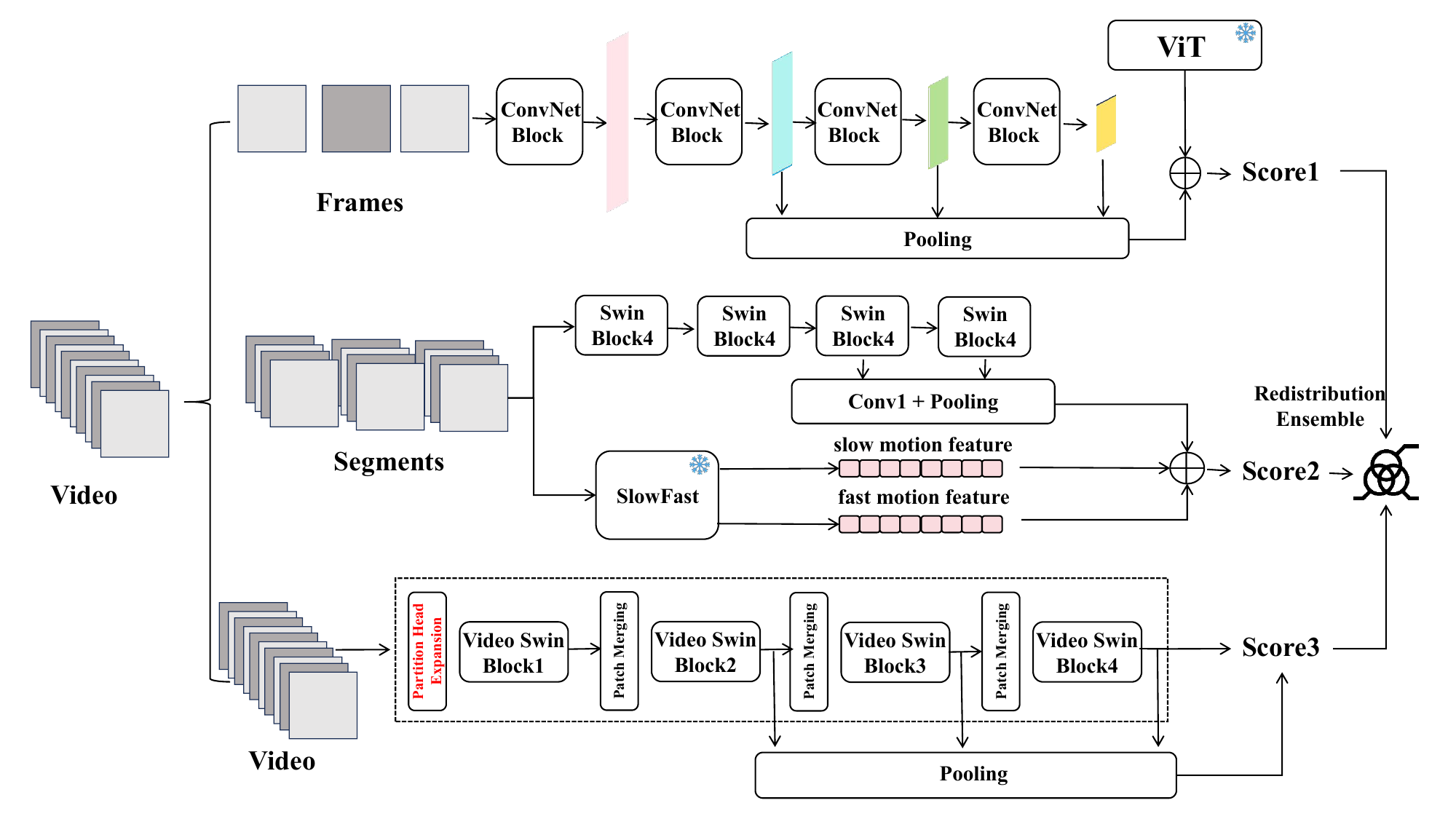}
    \caption{An overview of VideoFusion VQA framework, proposed by team VideoFusion.}
    \label{fig:videofusion}
\end{figure*}
This team proposes a three-level (frame-segment-video) integration framework~\cite{Xu2024VQA} for short-form UGC VQA in Fig.~\ref{fig:videofusion}. The contributions of this framework are summarized as follows: 
\begin{itemize}
    \item They propose a multi-level framework. Globally, a three-level architecture is proposed to capture features at each level, and locally, features on backbones from low level to high level are fused.
    \item Based on the view of data augmentation, data augmentation in spatial and temporal domains is employed on the three-level architecture respectively to improve the robustness of the model.
    \item In order to distinguish between two kinds of hard samples and relative rank information, they design an adaptive relative rank loss.
    \item By using the redistributed model integration strategy, the distribution of labels is aligned and the training of the model is more stable.
\end{itemize}

\noindent\textbf{Training Details.} They set the batch size to 4 and assigned a weight of 0.3 to the rank loss. They employ the AdamW optimizer with an initial learning rate of $1 \times 10^{-3}$. Following a warm-up period spanning 3 epochs, the learning rate is modulated using a cosine decay schedule. The weight decay for the optimizer is configured at $1 \times 10^{-2}$, and the model underwent training for a total of 30 epochs.

\noindent\textbf{Adaptive Rank-Aware Loss.}
In their method, they propose an Adaptive Rank-Aware Loss function to effectively handle the challenges posed by the coexistence of homogenous and heterogeneous data within Kwai UGC video dataset. This loss function is designed to differentiate between hard samples and enhance the model's ability to make fine-grained distinctions in video quality.

The loss function is formulated as follows:
\begin{align}
e(y_i^{gt}, y_j^{gt}) = \left\{\begin{array}{c}\quad1, \quad y_i^{gt} \geq y_j^{gt}\\
-1, \quad y_i^{gt} < y_j^{gt}
\end{array}\right.
\end{align}

\begin{align}
\mathcal{L}_{\text{r}} &= \frac{1}{m^2}\sum_{i=1}^{m}\sum_{j=1}^{m} \left[ \max(0, -e(y_i^{gt}, y_j^{gt}) (y_i-y_j) ) \right]^2 
\end{align}

\begin{align}
\mathcal{L}_{\text{Rank}} &= M \mathcal{L}_{\text{r}} + \lambda(1 - M)M_\alpha\mathcal{L}_{\text{r}}
\end{align}

In this equation, $M_\alpha$ denotes a distance indicator function that utilizes the threshold of the ground truth score of video pair $(y_i^{gt}, y_j^{gt}$, where $M_\alpha$ is 1 if ${\left| y_i^{gt}-y_j^{gt}\right|} < c$ else is 0 if${\left| y_i^{gt}-y_j^{gt}\right|} \geq c$. $m$ is the number of all pairs in the same batch, and $M = [y_i^{class} = y_j^{class}]$ is an indicator that is 1 if the videos in the pair have sample class label (homogeneous videos), and 0 otherwise. The margin parameter $\lambda$ is introduced to control the trade-off between homogeneous loss and non-homogeneous pairs loss.

\subsection{MC$^2$Lab}
This team utilizes the Swin Transformer V2~\cite{liu2022swinSwinV2} to extract features from salient regions within videos, employs ConvNext~\cite{liu2022convnetConvNetXt} for
capturing the overall distortion information of the videos, and leverages
the SlowFast~\cite{simpleVQA} network for extracting information related to motion distortions. These features are then concatenated. Ultimately, a three-layer
Multilayer Perceptron (MLP) is applied to yield the final quality score.

\noindent\textbf{Training Details.} 
This team utilizes the optimizer and schedule of FastVQA~\cite{wu2022fastFastVQA}.
The total epochs of the training procedure are 20. The batch size is 8, and the video clip length is 5. The input resolution of the video clip is 768×448 for
ConvNext~\cite{liu2022convnetConvNetXt} and 256×256 for Swin Transformer~\cite{liu2022swinSwinV2}. 

\noindent\textbf{Testing Details.}  For each video, They predict the quality score 5 times and average them as the final score.

\subsection{Padding}
This team proposes the zero-padding strategy~\cite{alrasheedi2023padding,liu2022partialpadding,hashemi2019enlargingpadding,dwarampudi2019effectspadding,naseri2023novelPadding} for S-UGC VQA, which is utilized in all scenarios requiring frame supplementation or duplication as:
\begin{itemize}
    \item When the number of frames read is less than the length of the video,
the remaining positions are now padded with tensors consisting entirely
of zeros.
\item During the processing of each video segment, if the length of the segment
is inadequate, it is now padded with a tensor that is entirely zeros.
\item If the actual count of video segments falls short of the minimum required number of video clips, tensors filled with zeros, equal in size to the first segment, are now appended until the minimum segment count is met.
\end{itemize}

There are two advantages of zero-padding for S-UGC VQA: (i) Zero-element padding does not introduce any extra information to the
image, thus avoiding the introduction of potential biases during the learning process. In contrast, other padding methods, such as replicating edge pixels, could introduce information that does not belong to the original video content, potentially affecting the network’s learning. (ii) Zero-padding helps maintain the spatial location of features within a
certain degree. Replicating the final frame suggests its multiple occurrences, which is detrimental to the convolutional network’s capacity to
learn features in relation to their positional context. This team has found that when they modified the approach to generate feature
inputs by substituting the prior method of replicating the final frame
with a strategy of zero-element padding, the resultant accuracy of the test set rose to 86.90\% from 86.16\%, thereby affirming their belief that padding with
zero elements is indeed a highly effective strategy.  

\subsection{ysy0129}
\begin{figure*}[htp]
\centering
\includegraphics[width=0.85\textwidth]{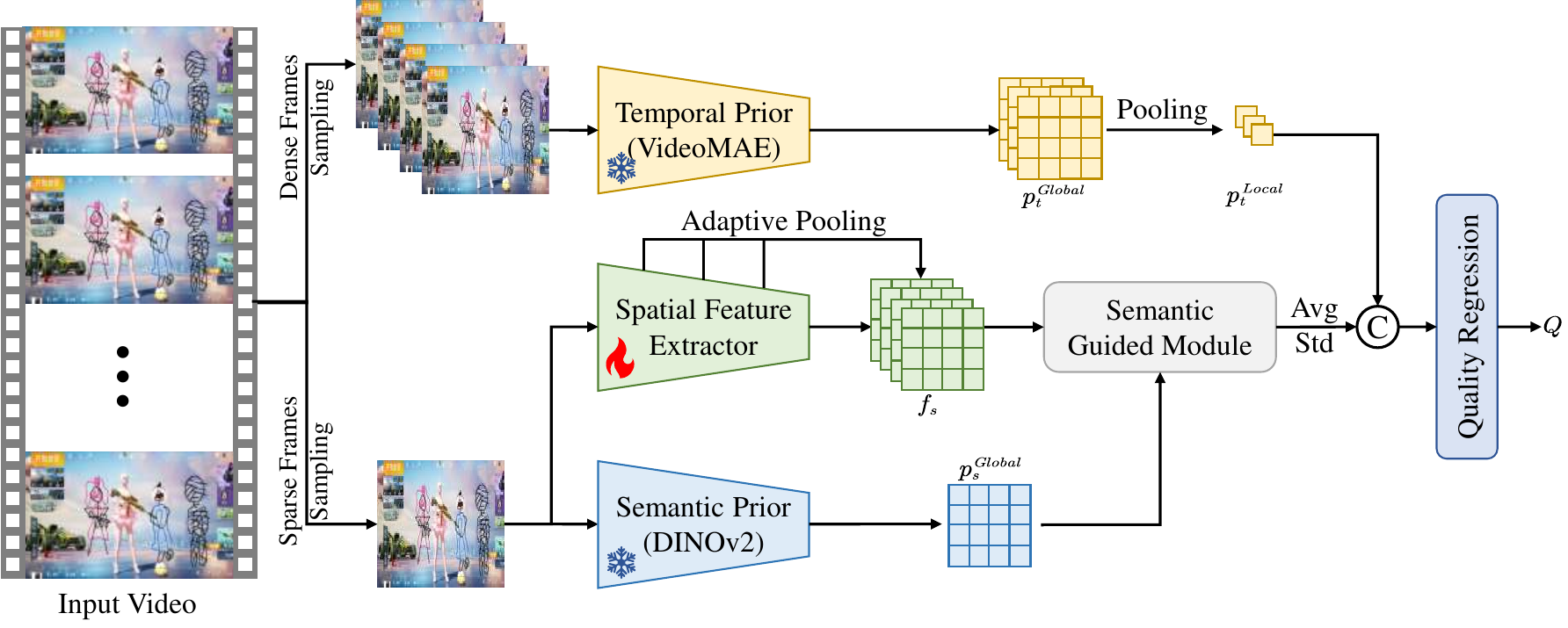}
\caption{The network architecture of the solution proposed by team ysy0129.}		
\label{fig:framework_ysy0129}	
\end{figure*}

\begin{figure}[htp]
\centering
\includegraphics[width=1.0\linewidth]{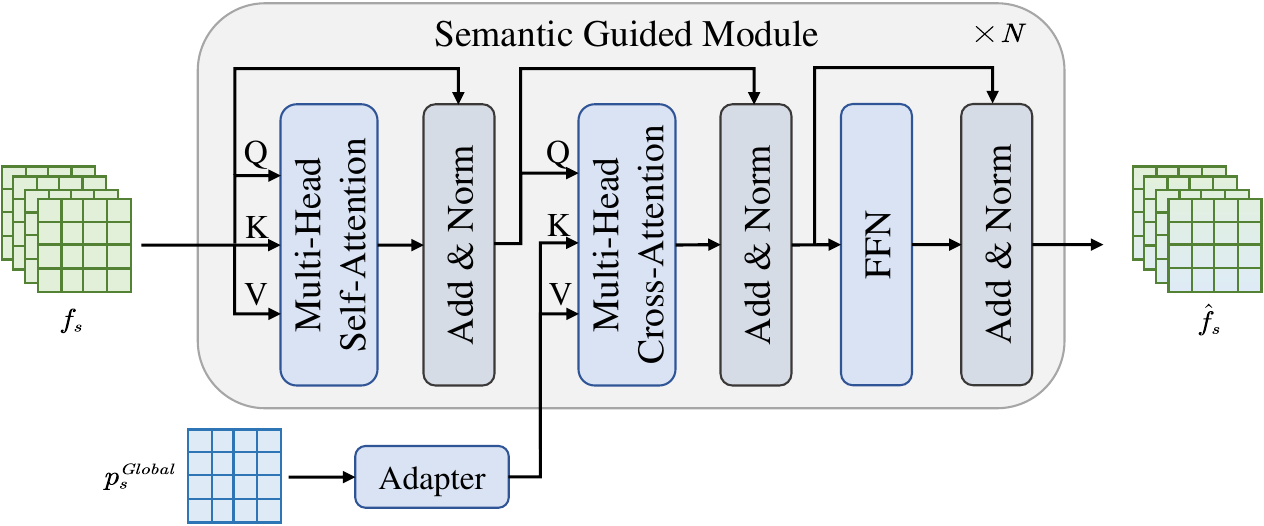}
\caption{The Semantic Guided Module of the solution proposed by team ysy0129.}		
\label{fig:crossattn_ysy0129}	
\end{figure}

\begin{figure}[htp]
\centering
\includegraphics[width=0.9\linewidth]{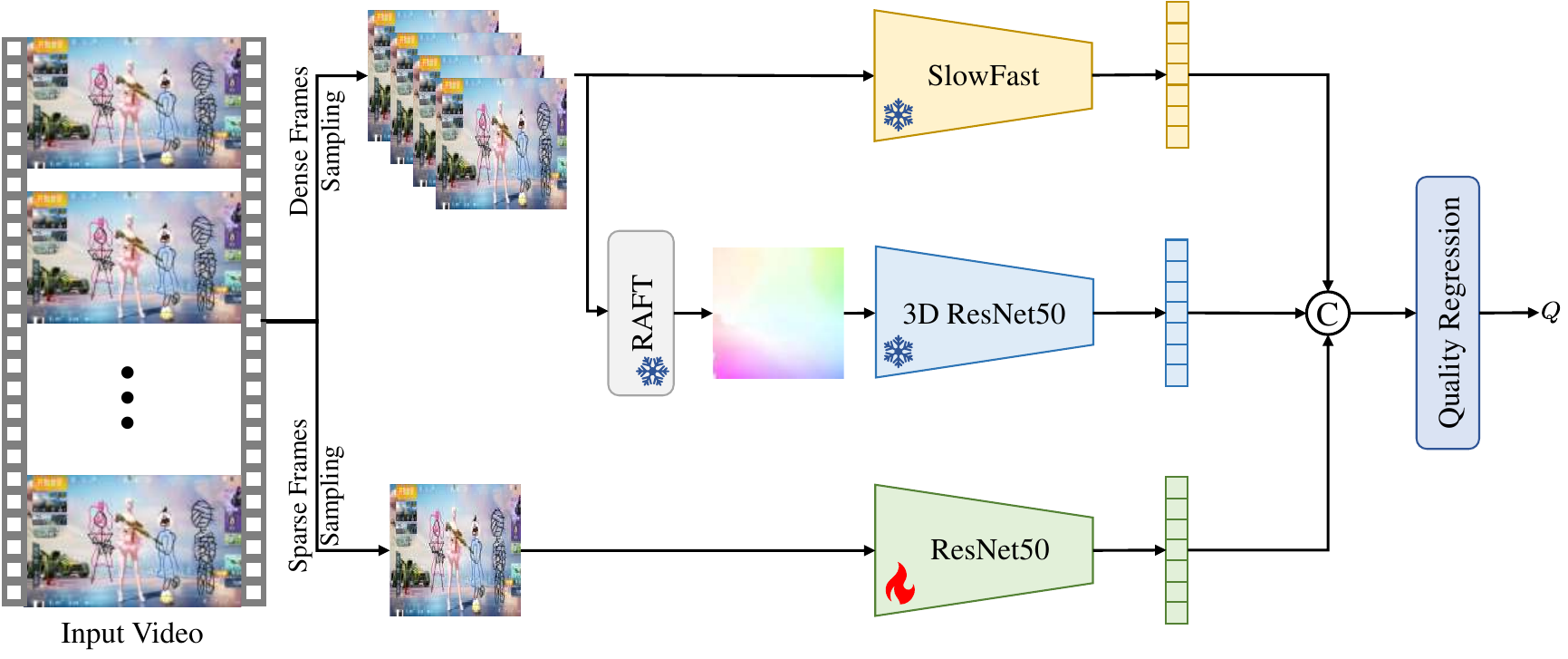}
\caption{The framework of SimpleVQA-OPT proposed by team ysy0129.}		
\label{fig:optical_flow_ysy1029}	
\end{figure}

This team proposes a video quality assessment method that embeds powerful self-supervised model priors. The framework of the proposed model is shown
in Fig.~\ref{fig:framework_ysy0129}. They sample frames from input video in sparse and dense manners, respectively, for extracting motion distortion and spatial distortion.

As for motion distortion, a self-supervised pre-trained model VideoMAE~\cite{tong2022videomae}
is utilized to extract motion features. When evaluating the quality of a
video, there exists a certain continuity in the quality between frames. Therefore, there is a certain amount of redundant information. Since the MAE
model learns the feature representation of the data through the reconstruction task of an autoencoder. In this process, the model might ignore
or compress redundant information. Consequently, Masked Autoencoders
(MAE) based model VideoMAE~\cite{tong2022videomae} is chosen as the video sequence extraction algorithm, which is related to the information redundancy.

As for spatial distortion, the pre-trained semantic prior model is adopted
to guide the spatial features. DINOv2~\cite{oquab2023dinov2} has been revealed powerful
semantic perception capability, attributed to self-supervised training on
a dedicated, diverse, and curated image dataset, which can facilitate our
model to focus on the semantic information which is related to the quality
of a UGC video.

Specifically, as shown in the framework, a spatial feature extractor with the backbone ConvNeXt~\cite{liu2022convnetConvNetXt} is adopted to capture the multi-scale spatial
distortion for video frames. Since the extracted features have different sizes, the extracted features are adaptive pooling to the same size and concatenated together as $f_s$.
Then, $f_s$ is enhanced by the semantic feature $p_s^\text{Global}$ generated by DINOv2~\cite{oquab2023dinov2} in the Semantic Guided Module (SGM).
As shown in Fig.~\ref{fig:crossattn_ysy0129}, the SGM consists of $N$ cross attention structure and an adapter composed of Linear-Relu-Linear architecture, which takes $f_s$ and $p_s^\text{Global}$ as input and output the guided feature $\hat{f}_s$.
Next, they apply global average and stand deviation pooling operations on $\hat{f}_s$ and concatenate them together with the motion features $p_t^\text{Local}$ extracted by VideoMAE~\cite{tong2022videomae} after pooling.
Lastly, the above spatial and motion features are fused together with a simple regression head, integrating both spatial and temporal aspects of the video.

\noindent\textbf{Training Details.}
In the training phase, their end-to-end network is trained with AdamW optimizer for 50 epochs. The initial learning rate and weight decay are set to $3e^{-5}$ and 0.05 respectively. And batch size is set as 4.

They uniformly partition each input video into 8 chunks. Within the spatial feature extractor branch, they resize the original video to dimensions of 520 $\times$ 520 and randomly sample 1 frame from each chunk. Subsequently, they crop the obtained 8 frames to the dimension of 448 $\times$ 448.
For the temporal feature branch, they uniformly selected 32 frames from each chunk and resized them to the dimension of 224 $\times$ 224.

\noindent\textbf{Testing Details.}
The video processing procedure is the same as the training process, ensuring consistency in the handling of data across both stages.
They conduct tests on two models, \textit{i.e.}, their VQA framework and SimpleVQA-OPT as shown in Fig.~\ref{fig:framework_ysy0129} and Fig.~\ref{fig:optical_flow_ysy1029}, respectively. The SimpleVQA-OPT is achieved by incorporating optical flow information into SimpleVQA~\cite{simpleVQA}. Inspired by StableVQA~\cite{kou2023stablevqa}, they utilize the RAFT to extract the optical flow.  Then, they utilize the pre-trained 3D ResNet of StableVQA~\cite{kou2023stablevqa}, which is pre-trained on a video quality assessment dataset with diversely-shaky UGC videos, to extract optical flow features. Subsequently,  these features are combined with spatial and motion features. They perform 8 tests for each model, as the spatial feature branch samples different frames during each test process.
All the scores are averaged to obtain the final results.

\subsection{lizhibo}
This team first segments the video into continuous chunks, then employs SlowFast R50~\cite{feichtenhofer2019slowfast} as the backbone to extract motion features from a key frame, and utilizes ResNet50~\cite{resnet} as the backbone to extract spatial features from all frames. Subsequently, a Multi-Layer Perceptron (MLP) network maps quality-aware features to chunk-level quality scores, followed by a temporal average pooling strategy to obtain the final video quality.

\noindent\textbf{Training Details.} 
This team first removes distorted videos from the training set and uses the first 500 video data as the validation set. The total epochs of the training process is 50, the initial learning rate of the Adam optimizer is 3e-5, and the batch size is 8. During the training process, the Exponential Moving Average (EMA)~\cite{Klinker2011ExponentialMAEMA} strategy is employed to enhance the model’s stability and improve the generalization ability of the model.

\noindent\textbf{Testing Details.}  
During the testing phase, they use the Exponential Moving Average (EMA) model for inference, reducing the risk of overfitting.

\subsection{YongWu}
This team utilizes various data augmentation for S-UGC VQA. At the onset of the competition, the team delves into various Video Quality Assessment (VQA) methodologies, including SimpleVQA~\cite{simpleVQA}, FastVQA~\cite{wu2022fastFastVQA}, and DOVER~\cite{Dover}. Following a thorough comparison, it is found that DOVER outperforms the others, aligning with the findings presented in KVQ~\cite{lu2024kvq}. Consequently, DOVER is chosen as the baseline model for the competition.

To improve the performance of the baseline model, they aim to acquire additional videos without compromising the quality assessment score of the training data. To achieve this, they employ various methods as follow:

\begin{itemize}

\item \textbf{Padding.} To prevent distortion caused by resizing, they implement zero-padding to preprocess the training and validation videos while maintaining the integrity of the quality assessment scores. Upon evaluation, this method does not yield favorable results for the baseline model. The reason for this is that the aesthetic branch of the baseline model relies on proportional scaling, and the technical branch samples fragments that are unaffected by resizing. However, this approach proves to be effective for FastVQA.

\item \textbf{Flip.} To maintain the quality assessment score, flipping videos with minimal text proved to be effective. Their analysis of the training data revealed that videos with little text remained unaffected in terms of quality scores post-flipping. They select 747 such videos and applied flipping to them.

\item \textbf{Average.} The dataset comprises an original video and five derivative versions, all classified under a single category. They organize the training data into separate folders, each containing videos of identical content. Subsequently, they randomly select two videos from each folder and calculate their average. Empirically, the quality score of the averaged video typically lies between that of the lower and higher-scored videos. For simplicity, they calculate the average quality score for these two videos, yielding a total of 418 new videos. Additionally, one could employ a predictive model to estimate the score and introduce slight variations to the averaged video's quality score. However, due to time constraints, no further exploration was undertaken. It is important to recognize that the video quality assessment model essentially functions to fit the quality scores, and having a greater number of sampling points enhances the accuracy of this fit.

\end{itemize}

\noindent\textbf{Rank Loss.} To boost the rank score, they have experimented with increasing the weight of rank loss from 0.3 to 0.5, which proved beneficial for the model. they also consider replacing the rank loss function:
\begin{align}
    L^{ij}_{rank} = max(0, |\hat{Q}_i - \hat{Q}_j| - sign(\hat{Q}_i, \hat{Q}_j) \cdot (Q_i - Q_j))
\end{align}

where $i$ and $j$ are two video indexes in a mini-batch, and $sign$ is formulated as:

\begin{align}
sign(\hat{Q}_i, \hat{Q}_j) = \left\{\begin{array}{c}\quad1, \quad \hat{Q}_i \geq \hat{Q}_j\\
-1, \quad \hat{Q}_i < \hat{Q}_j
\end{array}\right.
\end{align}

 but as the rank score had a negligible impact, this change does not significantly improve the model.

\noindent\textbf{Training Details.}
During the training phase, they train the end-to-end model using the AdamW optimizer for 50 epochs, with an initial learning rate of 0.001 and a weight decay of 0.05. The batch size was set at 12.

For training the model, they utilize 99 percent of the data instead of using cross-validation and normalized the original scores to the range of $[0, 1]$, aiding in model convergence. They sample 32 frames at 2-frame intervals from each video, with both the aesthetic and technical branches receiving inputs of size 224 $\times$ 224.

\noindent\textbf{Testing Details.} In the testing phase, they use the last saved epoch model with Exponential Moving Average (EMA)~\cite{Klinker2011ExponentialMAEMA} to enhance the model's stability and prevent overfitting.

\subsection{We are a team}
\begin{figure*}
    \centering
    \includegraphics[width=0.9\linewidth]{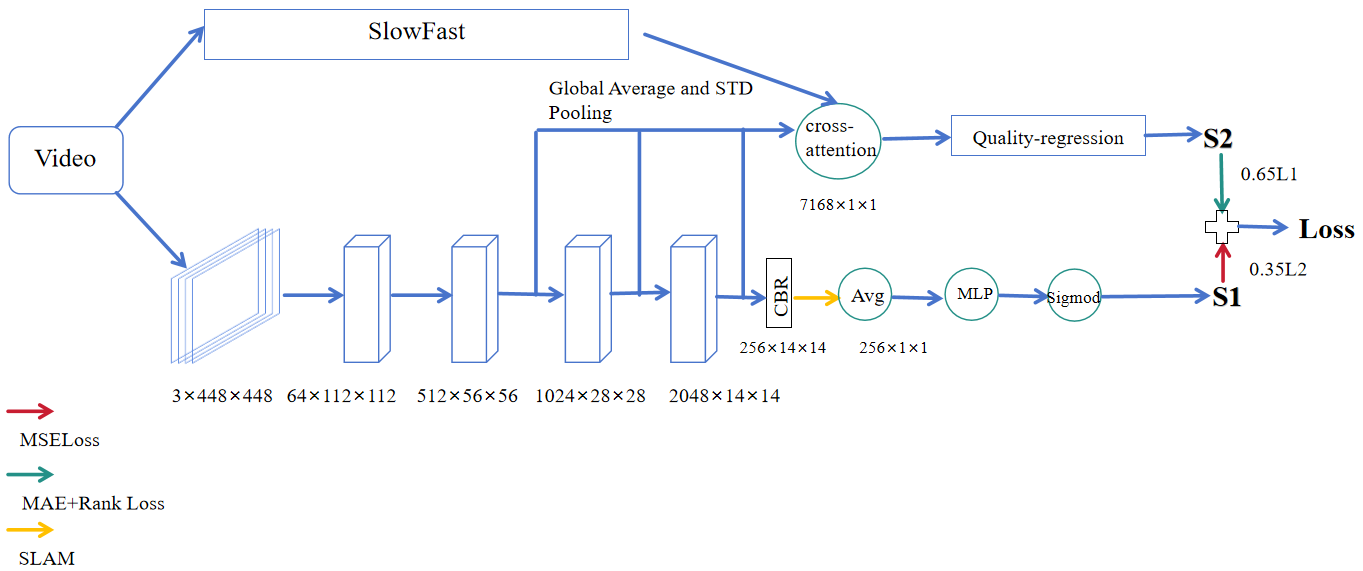}
    \caption{The network architecture of the solution proposed by team We are a team }
    \label{fig:we_are_a_team}
\end{figure*}

This team proposes a multi-task based short video quality assessment method~\cite{2024dulinaUGC-ICVQA}, which is shown in Fig.~\ref{fig:we_are_a_team}. Video quality assessment is closely related to the task of video complexity estimation. For example, when high-frequency noise is introduced into the video, the complexity increases, while when low-frequency blur is introduced, the complexity decreases. Therefore, this method introduces complexity pseudo-labels into the KVQ data set, uses a multi-task training method to evaluate the quality of short videos, and forces the backbone network to extract more general and robust features to meet the needs of two tasks at the same time. Their experimental results demonstrate that the performance of the video quality assessment task can also be improved when the video complexity estimation task is introduced. 

    Similar to most video tasks, their method also has two branches: spatial feature extraction branch and temporal feature extraction branch. For each video, they sample 1 frame per second and adjust its size to 520$\times$520, crop it to 448×448 and use ResNet50~\cite{resnet} for spatial feature extraction; sample 32 frames per second and adjust the size to 224×224 and use SlowFast~\cite{feichtenhofer2019slowfast}for temporal feature extraction. Use cross-attention on the extracted spatial and temporal features to obtain spatio-temporal fusion features. Finally, the quality assessment head performs video quality score regression based on spatio-temporal fusion features, and the complexity estimation head performs video complexity score regression based on spatial features.

\noindent\textbf{Training Details.}
In the training phase,  they use the Adam optimizer for 100 epochs of end-to-end training, with the initial learning rate and decay weight set to 3e-5 and 0.9, and the batch size set to 32. For the quality assessment task, MAE loss and Rank loss are used, and the complexity estimation task uses MSE loss.

\noindent\textbf{Testing Details.}
During the testing phase, for each video, the model only predicts the quality score once as the final result.

\subsection{dulan}
This team has explored different VQA methods such as FastVQA~\cite{wu2022fastFastVQA} and Dover~\cite{wu2023exploringDover}. The empirical results demonstrate that these previously proposed methods can provide satisfactory performance. However, they usually struggle to obtain optimal performance owing to the existence of domain shifts caused by different data distributions. To solve the above issue, 
this team designed a simple yet effective UGC VQA model based on  SimpleVQA~\cite{simpleVQA},  which trains an end-to-end spatial feature extractor to learn the quality-aware feature representation.

To enhance the performance of feature extraction, they adopt the SlowFast~\cite{feichtenhofer2019slowfast} approach, which is known for its effectiveness in capturing temporal and spatial features in videos. Following the approach in~\cite{wu2023video}, they utilize the Tiny Swin-Transformer~\cite{liu2021swin} as the backbone network, given its superior performance in various visual recognition tasks.
Furthermore, they incorporate the Exponential Moving Average (EMA)~\cite{Klinker2011ExponentialMAEMA} strategy to improve the model's stability during training, which in turn enhances its performance on the test set.

By leveraging the predictions of multiple models, the strengths of each model can be harnessed to generate a more resilient and precise overall prediction. During the inference stage, they implement an ensemble strategy that combines model parameters from different epochs. This ensemble technique not only diversifies the model's predictions but also significantly contributes to the overall improvement of its performance.
The integration of these techniques effectively aggregates multi-view information, thereby enhancing the capabilities of their UGC VQA model, positioning it as a robust and efficient solution for this VQA challenge.

\noindent\textbf{Training Details.}
Their model is based on SimpleVQA~\cite{simpleVQA} and they modify the spatial feature extraction module. Specifically, they utilize Swin Transformer~\cite{liu2021swin} pre-trained on ImageNet and SlowFast network~\cite{feichtenhofer2019slowfast} pre-trained on Kinetics as the feature extraction model. 
During the training phase, they employ the EMA strategy to enhance the stability of the model’s convergence, which will be elaborated on the experimental results of the EMA strategy in Table~\ref{tab:comparison_dulan}. There are 26 millions of parameters in the model. Batch size is set to 4, the learning rate is set to \(10^{-4}\), and the AdamW optimizer is adopted with a weight decay of $5\times10^{-4}$. In addition, they use the cosine decay learning rate with the minimum learning rate of \(10^{-7}\) and use linear preheating in the first 2 epochs with start learning rate $5\times10^{-7}$.

\noindent\textbf{Testing Details.}
In the test stage, they adjust the model’s weights by assigning a bigger weight to the model with the lowest loss and smaller weights to the models from the subsequent epoch, thereby achieving a fusion of model parameters. 

\noindent\textbf{Analysis.}
(1) \textbf{EMA strategy.} They employ the EMA strategy and preserve models both with and without the application of EMA. Table~\ref{tab:comparison_dulan} presents a comparison of the performance between these two sets of models. In this comparison, ``Model-EMA'' refers to the model that employs the EMA strategy, whereas ``Model-None'' does not utilize this strategy. It is evident from the table that the application of the EMA strategy significantly enhances the performance of the model.

\begin{table}[htbp]
\centering
\caption{Validation Accuracy Comparison between Model-EMA and Model-Without EMA. $\downarrow$ donates the lower, the better and $\uparrow$ donates the higher, the better. The experimental results are from the team dulan.}
\label{tab:comparison_dulan}
\begin{tabular}{lcc}
\toprule
\textbf{Metric} & \textbf{Model-EMA} & \textbf{Model-None} \\
\midrule
SRCC$\uparrow$ & \textbf{0.8554} & 0.3776 \\
PLCC$\uparrow$ & \textbf{0.8598} & 0.3979 \\
KRCC$\uparrow$ & \textbf{0.6646} & 0.2565 \\
RMSE$\downarrow$ & \textbf{0.3220} & 0.6672 \\
\bottomrule
\end{tabular}
\end{table}

\textbf{(2) Truncation.} In addition, since the optimal range for the output scores is between 1 and 5, they implemented data truncation for unreasonable outputs, setting negative values to 0. However, this measure inadvertently leads to a decrease in the output scores. 

(3) \textbf{Ensembles and fusion strategies.} This team argues that no single model can capture all aspects of the data distribution. By combining the predictions of multiple models, the strengths of each model can be leveraged to produce a more robust and accurate overall prediction. They perform a weighted sum fusion based on models saved at different epochs. In the training process, model\text{-}l represents the model weights with the lowest loss value achieved during the training stage. Additionally, model\text{-}s refers to the model weights saved from the previous epoch. The weights \(p_1\) and \(p_2\) are the proportions of the output results from model\text{-}l and model\text{-}s, respectively. The final score is calculated as follows: 
\begin{equation}
    Score=p_{1}\times W_{model\text{-} s}+p_{2}\times W_{model\text{-} l}
\end{equation}
Through experiments, they identify the optimal ensemble weight ratio yielding the highest score. The model with the lowest loss is assigned a weight of 0.6, and the model from the subsequent epoch is assigned a weight of 0.4.

\subsection{D-H}
This team utilizes SimpleVQA~\cite{simpleVQA} model. The S-UGC videos are first sampled and each frame is divided into patches, which are inputted to the spatio-temporal features models and regression module to obtain the predicted score.  The loss function is the summation of PLCC loss and ranking loss.

\noindent\textbf{Training Details.}
They randomly sampled 20\% samples from each fraction segment of training samples as a validation set, and the
remaining samples were used as the training set. The model weights with the highest evaluation index of the validation set are stored in each epoch. They also utilize the exponential moving average strategy to make the training to be stable and smooth.

% \input{sec/5_experiment}

% \section{Conclusion}

% ~\cite{li2024ntire}
% \clearpage
\section*{Acknowledgements}
This work was partially supported by the Humboldt Foundation. We thank the NTIRE 2024 sponsors: Meta Reality Labs, OPPO, Kuaishou, Huawei and University of W\"urzburg (Computer Vision Lab). We also thank Kuaishou for sponsoring this challenge.

\appendix
\section{Teams and Affiliations}
\subsection*{NTIRE2024 Organizers}

\noindent\textit{\textbf{Title:}} NTIRE 2024 Challenge on Short-form UGC Video Quality Assessment

\noindent\textit{\textbf{Members:}}
Xin Li\textsuperscript{1} (\textcolor{magenta}{lixin666@mail.ustc.edu.cn}), Kun Yuan\textsuperscript{2}, Yajing Pei\textsuperscript{1}, Yiting Lu\textsuperscript{1}, Ming Sun\textsuperscript{2}, Chao Zhou\textsuperscript{2}, Zhibo Chen\textsuperscript{1}, and Radu Timofte\textsuperscript{3} 

\noindent\textit{\textbf{Affiliations:}}

\noindent \textsuperscript{1} University of Science and Technology of China

\noindent \textsuperscript{2} Kuaishou Technology

\noindent \textsuperscript{3} University of W\"urzburg

\subsection*{SJTU MMLab}

\noindent\textit{\textbf{Title:}} Enhancing Blind Video Quality Assessment with Rich Quality-aware Features

\noindent\textit{\textbf{Members:}}

\noindent Wei Sun\textsuperscript{1} (\textcolor{magenta}{sunguwei@sjtu.edu.cn}), Haoning Wu\textsuperscript{2}, Zicheng Zhang\textsuperscript{1}, Jun Jia\textsuperscript{1}, Zhichao Zhang\textsuperscript{1}, Linha Cao\textsuperscript{1}, Qiubo Chen\textsuperscript{2}, Xiongkuo Min\textsuperscript{1}, Weisi Lin\textsuperscript{2}, and Guangtao Zhai\textsuperscript{1}.

\noindent\textit{\textbf{Affiliations:}}

\noindent \textsuperscript{1} Shanghai Jiao Tong University

\noindent \textsuperscript{2} Nanyang Technological University 

\noindent \textsuperscript{3} Xiaohongshu

\subsection*{IH-VQA}

\noindent\textit{\textbf{Title:}} Ensemble-based Video Quality Assessment System

\noindent\textit{\textbf{Members:}}

\noindent Jianhui Sun (\textcolor{magenta}{55530544@qq.com}), Tianyi Wang, Lei Li, Han Kong, Wenxuan Wang, Bing Li and Cheng Luo

\noindent\textit{\textbf{Affiliations:}}

\noindent WeChat

\subsection*{TVQE}

\noindent\textit{\textbf{Title:}} Tencent Video Quality Evaluator

\noindent\textit{\textbf{Members:}}

\noindent Haiqiang Wang\textsuperscript{1} (\textcolor{magenta}{walleewang@tencent.com}), Xiangguang Chen\textsuperscript{1}, Wenhui Meng\textsuperscript{1}, Xiang Pan\textsuperscript{1}, Huiying Shi\textsuperscript{2}, Han Zhu\textsuperscript{2}, Xiaozhong Xu\textsuperscript{1}, Lei Sun\textsuperscript{1}, Zhenzhong Chen\textsuperscript{2}, and Shan Liu\textsuperscript{1}

\noindent\textit{\textbf{Affiliations:}}

\noindent \textsuperscript{1} Tencent

\noindent \textsuperscript{2} Wuhan University

\subsection*{BDVQAGroup}

\noindent\textit{\textbf{Title:}} UGC Video Quality Assessment Method based on LMMs

\noindent\textit{\textbf{Members:}}

\noindent Fangyuan Kong (\textcolor{magenta}{kongfangyuan@bytedance.com}), Haotian Fan, Yifang Xu

\noindent\textit{\textbf{Affiliations:}}

\noindent ByteDance, Shenzhen

\subsection*{VideoFusion}

\noindent\textit{\textbf{Title:}} VideoFusion

\noindent\textit{\textbf{Members:}}

\noindent Haoran Xu (\textcolor{magenta}{ xhr964691257@163.com}), Mengduo Yang, Jie Zhou, and Jiaze Li

\noindent\textit{\textbf{Affiliations:}}

\noindent Zhejiang University

\subsection*{MC$^2$Lab}

\noindent\textit{\textbf{Title:}}  Short-form UGC Video Quality Assessment: A
Combined Approach

\noindent\textit{\textbf{Members:}}

\noindent Shijie Wen (\textcolor{magenta}{wenshijie@buaa.edu.cn}) and Mai Xu

\noindent\textit{\textbf{Affiliations:}}

\noindent Department of Electronic and Information Engineering, Beihang University, Beijing, China

\subsection*{Padding}

\noindent\textit{\textbf{Title:}}  How Should We Apply Padding?

\noindent\textit{\textbf{Members:}}

\noindent Da Li (\textcolor{magenta}{1048222617@qq.com}) 

\noindent\textit{\textbf{Affiliations:}}

\noindent Southern University of Science and Technology

\subsection*{ysy0129}

\noindent\textit{\textbf{Title:}}  Video Quality Assessment Based on Self-Supervised
Model Priors

\noindent\textit{\textbf{Members:}}

\noindent Shunyu Yao (\textcolor{magenta}{shunyuyao19@gmail.com}), Jiazhi Du and Wangmeng Zuo

\noindent\textit{\textbf{Affiliations:}}

\noindent Harbin Institute of Technology

\section*{lizhibo}

\noindent\textit{\textbf{Title:}}  Short-form Video Quality Evaluator

\noindent\textit{\textbf{Members:}}

\noindent Zhibo Li (\textcolor{magenta}{lizhibo\_zyber@bupt.edu.cn}), Shuai He, Anlong Ming, Huiyuan Fu, Huadong Ma

\noindent\textit{\textbf{Affiliations:}}

\noindent Beijing University of Posts and Telecommunications, China

\section*{YongWu}

\noindent\textit{\textbf{Title:}}  Data augmentation for the Short-form UGC Video Quality Assessment

\noindent\textit{\textbf{Members:}}

\noindent Yong Wu (\textcolor{magenta}{wuyong139@hotmail.com}), Fei Xue and Guozhi Zhao

\noindent\textit{\textbf{Affiliations:}}

\noindent China Merchants Bank

\section*{We are a team}

\noindent\textit{\textbf{Title:}}  Short video quality assessment combined with video complexity

\noindent\textit{\textbf{Members:}}

\noindent Lina Du(1340398672@qq.com),Jie Guo,Yu Zhang,Huimin Zheng,Junhao Chen and Yue Liu

\noindent\textit{\textbf{Affiliations:}}

\noindent Shandong Jianzhu University

\section*{dulan}

\noindent\textit{\textbf{Title: No title}}

\noindent\textit{\textbf{Members:}}

\noindent Dulan Zhou (\textcolor{magenta}{zdl190301000@163.com}), Kele Xu, Qisheng Xu and Tao Sun

\noindent\textit{\textbf{Affiliations:}}

\noindent Key Laboratory for Parallel and Distributed Processing,
Changsha, China

\section*{D-H}

\noindent\textit{\textbf{Title: No title}}

\noindent\textit{\textbf{Members:}}

\noindent Zhixiang Ding (\textcolor{magenta}{dingzhixiang23@mails.ucas.ac.cn}) and Yuhang Hu

\noindent\textit{\textbf{Affiliations:}}

\noindent \textsuperscript{1} Institute of Automation, CAS

\noindent \textsuperscript{2} University of Chinese Academy of Sciences

{
    \small
    \bibliographystyle{ieeenat_fullname}
    \bibliography{main}
}


\end{document}